%% file: photon.tex
\documentclass[print]{aa}
\usepackage{amsmath}
\usepackage[pdftex]{graphicx}
\usepackage{pstricks}
\usepackage{array}
\usepackage{longtable}
\usepackage[utf8]{inputenc}
\usepackage{subfigure}
\usepackage{sidecap}
\usepackage{natbib}
\usepackage{txfonts}

\begin{document}

\title{Bayesian Photon Counting with EMCCDs}

   \author{Kennet B.W. Harps\o e\inst{1,2}, Michael I. Andersen\inst{1} \and Per Kj\ae gaard\inst{1}}
   
   \institute{Niels Bohr Institute, University of Copenhagen,
              Juliane Maries Vej 30, 2100 Copenhagen \O, Denmark\\
              \email{harpsoe@nbi.ku.dk}
           \and
              Centre for Star and Planet Formation, Natural History Museum of Denmark, University of Copenhagen, \O ster Voldgade 5-7, 1350 Copenhagen K, Denmark\\
           }


 
  \abstract
   {The EMCCD is a CCD type that delivers fast readout and negligible detector noise, making it an ideal detector for high frame rate applications. Because of the very low detector noise, this detector can potentially count single photons.}
   {Considering that an EMCCD has a limited dynamical range and negligible detector noise, one would typically apply an EMCCD in such a way that multiple images of the same object are available, for instance, in so called lucky imaging. The problem of counting photons can then conveniently be viewed as statistical inference of flux or photon rates, based on a stack of images.}
   {A simple probabilistic model for the output of an EMCCD is developed. Based on this model and the prior knowledge that photons are Poisson distributed, we derive two methods for estimating the most probable flux per pixel, one based on thresholding, and another based on full Bayesian inference.}
   {We find that it is indeed possible to derive such expressions, and tests of these methods show that estimating fluxes with only shot noise is possible, up to fluxes of about one photon per pixel per readout.}
   {}

   \maketitle

\input{EMCCDPrincible2}
\input{Strategies}

\input{Results}

\appendix
\input{EMteori}

\bibliographystyle{aa}
\bibliography{photon}

\end{document}

%% file: EMCCDPrincible2.tex
\section{Introduction}
As a detector for light in the visible, UV, and X-ray parts of the electromagnetic spectrum, the CCD is ubiquitous in astronomy. CCDs have a high quantum efficiency and low dark current when operated at temperatures of $\approx$ $-100^{\circ} C$. Under optimal conditions the dominant source of noise in the CCD itself is the readout noise. This is the noise in the readout amplifier, which converts the number of photoelectrons in each pixel to a digital ADU (Analog to Digital Unit) value. In the best CCD, under optimal conditions and very slow readout speeds ($\approx 10^{5}$ pixels per second), the noise of the conversion can be as low as 2-3 electrons. With higher readout speeds this performance is worse, with noise in the tens or even hundreds of electrons per readout.

\begin{figure}
	\begin{center}
	\includegraphics[width=0.8\columnwidth]{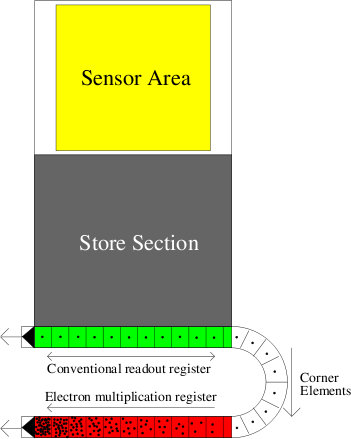}
	\end{center}
	\caption{Schematic drawing of a frame transfer EMCCD. After exposure, the image is rapidly parallely shifted into the storage area, where it is read out via the green serial register. The red register is an extension of the serial register where the pixels are shifted with significantly higher voltages, hereby generating cascade amplification via impact ionisation.}
	\label{fig:EMCCD}
\end{figure}

However, even with the very lowest achievable readout noise of a conventional CCD, it is still impossible to count photons (photoelectrons). At low photon rates, of about one photon per readout, the signal will be washed out by readout noise. One possible way out of this problem is the Electron Multiplying CCD (EMCCD), also known under the E2V trade name L3CCD. { An EMCCD is identical to a conventional CCD in terms of qantum efficiency, dark current etc., but its serial readout register has been extended as in Fig. \ref{fig:EMCCD}.} In the extended part of the register, the voltage  used to shift the captured electrons from pixel to pixel is not in the normal $5V$ range, but rather, it has been increased to around $40V$. This implies that the probability that an electron will knock another electron out of a bound state has been dramatically increased. Such an event is known as an impact ionisation event. The original electron will not be consumed but will generate a new electron hole pair. The event will thus effectively multiply the electron, similarly to the dynodes of a photomultiplier tube. The generated electron will subsequently be caught in the pixel ready to be shifted, together with the original electron, to the next pixels, where more impact events may be generated.
The probability of an impact ionisation per shift per electron is low, on the order of $p_m \approx 0.01$. But if the number of stages in the extended register is large, an amplification of two to three orders of magnitude can be achieved \citep{Hynecek2003}. 

Due to the stochastic nature of the impact ionisation events, the number of electrons in a cascade from one photo-electron is not constant; that is, the gain of the electron multiplying register is random, in a similar way to an avalanche diode or even a Geiger counter. This leads to a number of complications that will be addressed in later sections; however, the EMCCD has the potential to count single photons.

EMCCDs are typically manufactured as frame transfer CCDs. In a frame transfer CCD, half of the pixels are covered by a thin metal film. Once an image has been exposed, it is then shifted to the area covered by this film, from where it is read out while the next image is exposed. This method optimizes the duty cycle, as the time it takes to shift the image is much shorter than the time it takes to read out the image. So, as long as the exposure time is longer than the readout time for the image, essentially all photons reaching the detector are utilised.

We foresee two areas in astronomy where the photon counting capability is potentially very useful: namely, high framerate imaging--- for instance, { lucky imaging \citep{Tubbs2003,Fried78} and speckle imaging \citep{Baldwin1986}}--- and faint object spectroscopy.

In high framerate applications the individual frames will usually be severely photon-starved due to the shot exposure times, and the applicability of photon counting is therefore obvious.    

Faint objects pose a special problem in spectroscopy, notably in the red part of the spectrum, where night sky lines dominate the background. In the presence of night sky emission lines, spectra are best recorded at a moderately high resolution, at which most of the spectral bins are unaffected by night sky lines. For the night sky emission spectrum of the Earth atmosphere this implies a resolution of 5000--10000. As an example, a 2-4m class telescope is large enough to collect a sufficient number of photons from a 21--22 magnitude supernovae, to provide a R$\approx$500 spectrum for classification. However, in order to eliminate the impact of the sky lines, it is to be recorded at resolutions of 5000--10000, and subsequently rebinned, ignoring spectral channels affected by sky lines. With conventional CCDs, such an observation will for any reasonable exposure time be completely detector noise limited, and hence not practical. 

This problem can be solved by the use of photon counting with an EMCCD. This would make it possible to optimally utilize these intermediate sized telescopes for faint object spectroscopy. In the near future, large scale surveys, such as Pan-STARRS \citep{Kaiser} and LSST \citep{Strauss}, will turn up thousands of transient and/or rare objects, which will require spectroscopic follow-up. There will simply not be enough 10~meter class telescope time available for this. It is therefore important that some 2--4~meter class telescopes are equipped with optimized faint object spectrographs, such that part of the demand for spectroscopic followup from these surveys can be addressed with these smaller telescopes. Other authors have entertained the idea of photon counting spectrographs, mostly for time resolved spectroscopy, see for instance \cite{Tulloch2011}.

Another effect of such a spectrograph would be ``online'' access to the spectrum being recorded, making it possible to interrupt the integration when a certain desired signal-to-noise ratio has been reached, making more efficient use of telescope time possible.

The advent of the EMCCD has inspired other authors to develop methods for photon counting, based on thresholding and multithresholding \citep{Lantz2008,Basden2004a,Basden2003}.
Considering that an EMCCD has a limited dynamical range and negligible readout noise, one would typically apply an EMCCD in such a way that multiple images of the same object are available; as, for instance, in lucky imaging. The problem of counting photons can then conveniently be viewed as statistical inference of flux or photon rates, based on a stack of images.

We will here develop two intuitive methods for estimating photon rates with only shot noise (at low fluxes), which is essentially the same as counting photons: one based on simple thresholding, and one based on Bayesian inference.

The thresholding method is computationally simple and thus very fast, but one of the assumptions in its derivation is that the photon rate is constant from image to image in the stack. This is a reasonable assumption in faint object spectroscopy, but in high framerate imaging it is not strictly true because of seeing fluctuations. 

The other method based on full Bayesian inference can theoretically handle fluctuating photon rates, but at the cost of much more computational effort.  

%% file: Strategies.tex
\section{Description of EMCCD Data}

In the literature it is agreed upon that the multiplication of an ideal electron multiplier can be described as an exponential distribution in which the scale is the multiplication gain, see \cite{Basden2004,Lantz2008}. A distribution for the output can be derived via probability generating functions; see \cite{Matsuo1985,Basden2004} claim that this expression reduces to an exponential distribution when the probability of multiplication is low in every step in the multiplier, but no proof is provided to this claim. In fact, it has not been possible to find such a proof in the literature. Nevertheless, the exponential distribution of output is widely assumed in papers on EMCCDs, { and the claim appears to be supported empirically by data as well, for instance in Fig. \ref{fig:fit}, where a histogram of EMCCD output forms a straight line in a log plot.

A simulation of the EMCCD cascade amplification was made, in order to investigate if the distribution function is exactly exponential, and if not, how much it deviates. It was found that for small output values, it is even not monotonically decreasing, but has a maximum around output values of a few. The relative deviation from an exponential distribution is of the order of a few percent.}

\subsection{The Cascade Amplification Process}

According to \cite{Basden2004}, if we have a register where there is a probability $p_m$ that the one electron is multiplied in one stage of the multiplier, then the multiplication gain of $m$ steps is given by
\begin{equation}
\gamma = (1 + p_{\mu})^m
\end{equation}
Now let $X$ be the stochastic variable describing the number of cascade electrons resulting from one electron through the EM amplifier. Then the Probability Density Function (PDF) of $X$ is given by  
\begin{equation}
P(X=x) = \frac{1}{\gamma} e^{-x/\gamma } H(x)
\label{exponentialphoton}
\end{equation}
where $H$ is a Heaviside function, which is the integral of the delta function, and $\gamma$ is the EM amplification. { This distribution is continuous, even though the number of output electrons is formally a discrete number, but the number of output electrons after the amplification is high and it is common practice to go to the continuous limit. It can therefore conveniently be handled in the continuous limit.}

To get the distribution for two input electrons, we assume that the two electrons are multiplied independently. We can think of the size of the avalanche of electrons we get from the multiplication of one electron as a stochastic variable $X_1$. Similarly we will think of the avalanche from the other electrons as the stochastic variable $X_2$. The total size of the avalanche will then be $Y = X_1 + X_2$. From statistics we know that the distribution of such a sum is the convolution of the distributions of the individual variables.
\begin{align}
P(X_1+X_2=x) & =  \int_{0}^{x} P(X_1=t)P(X_2=x-t)dt \\
            & =  \int_{0}^{x} \frac{e^{-t/\gamma}}{\gamma} \frac{e^{-(x -t)/\gamma}}{\gamma} dt \\
            & =  \frac{1}{\gamma^2} \int_{0}^{x} e^{-x/\gamma} dt \\
            & =  \frac{x}{\gamma^2} e^{-x/\gamma}
\end{align}
{ where this integral is known as the convolution product.}
This can be repeated and the general result for $n$ convolutions, { corresponding to the case of $n$ electrons}, is known in the mathematical literature as the Erlang distribution, a special case of the more general gamma distribution. In the gamma distribution $n$ is not restricted to the integers.
\begin{equation}
\label{eq:ProbPlot}
P(Y=y \mid n)  = y^{n-1} \frac{e^{ -\frac{y}{\gamma} } }{\gamma^{n} \Gamma (n)}
\end{equation}
Here $y$ is the number of cascade output electrons and $n$ is the number of input electrons.
The gamma distribution has expectation $n\gamma$ and variance $n\gamma^2$. For large $n$ the distribution converges towards a Gaussian with expectation $n\gamma$ and variance $n\gamma^2$.
{ According to the data sheet for the CCD used for our investigations, the EM amplifier is linear; that is, $y \approx \gamma n$ up to about $4 \cdot 10^{5}$ output electrons, which at a typical gain of $\gamma \approx 1000$, corresponds to several hundred input photoelectrons. We therefore assume that the assumption of independent multiplication is valid for $n$ up to several hundred.}

\begin{figure}
\centering
\includegraphics[width=1.0\columnwidth]{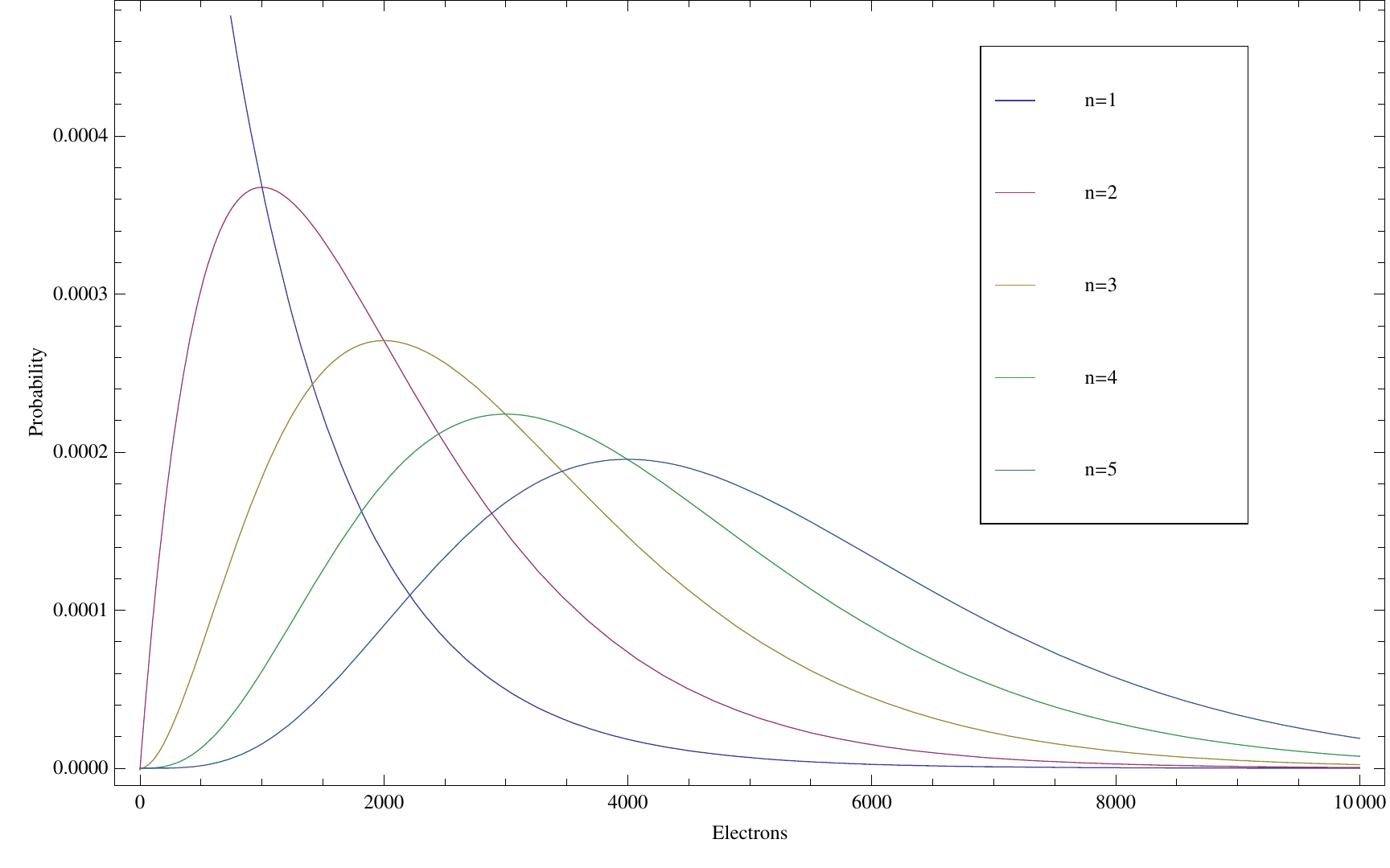}
\caption[Plot of Erlang distributions]{Plot of gamma distributions as defined in equation \ref{eq:ProbPlot}, for $n=\{1,2,3,4,5,6\}$ and $\gamma=1000$. $\gamma$ is a scale parameter, it will only scale the numbers on the $x$-axis, not change the functional form.}
\label{fig:ProbPlot}
\end{figure}

From Fig. \ref{fig:ProbPlot} it is clear that the distributions for different $n$ have a large overlap. So from just the readout we cannot reliably distinguish, say, one and three photoelectrons in the same pixel. 
This problem can be dealt with in different ways. One could try to keep the light levels so low that the probability of two electrons coinciding in the same pixel is close to zero, as we have done so far. 

\subsection{Simple Integration}

The mean value of the gamma distribution is $n\gamma$, as mentioned before, which means that we could just consider an EMCCD to be a device in which all the photoelectrons have been amplified with $\gamma$. That is, we could operate the EMCCD like any other CCD by exposing the CCD for a certain amount of time $t$, read it out, and then divide the result by $\gamma$. Unfortunately this approach will introduce additional noise on top of the photon shot noise.

If we let an average of $k$ photons hit a pixel in an EMCCD, and assume simple amplification with a factor of $\gamma$, the calculated signal to noise ratio, taking into consideration ordinary shot noise, will be
\begin{equation}
S/N=\frac{\gamma k}{\sqrt{\sigma^{2}_{shot}+\sigma^{2}_{EM}}}=\frac{\gamma k}{\sqrt{\gamma^2 k+\gamma^2 k}}=\frac{k}{\sqrt{2k}}
\end{equation}
Here we have used the normal scaling law for variance to scale the shot noise. We get 2 times the variance of conventional shot noise. This is usually known as the excess noise factor \citep{Hynecek2003}. 
An excess noise factor of one means that we have only shot noise; that is, we can count photons. Conversely an excess noise factor of two means that we do no better than the simple integration above.

This excess noise factor of two effectively lowers the QE (quantum efficiency) of the EMCCD, in the sense that we have to count twice as many photons to achieve a given S/N.
To get around this limitation we will have to embark on some kind of photon counting.

\subsection{Spurious Electrons}
In all CCDs, random spurious single electrons will arise in the parallel and serial registers, even without exposure to light, as an effect of the horizontal and vertical shift operations. These events are rare and on a single electron level, therefore they have not been reported from conventional CCDs as they will not be discernible in the presence of readout noise. But in an EMCCD these spurious electrons will be cascade-amplified in the EM stage, giving rise to a detectable signal.

{ We will assume that the event of a spurious electron is rare, integer, and uncorrelated; and therefore assume that they can be modeled with a Poisson distribution.}

In an EMCCD, spurious electrons can be generated in three distinct places: in the parallel register, in the serial register, or in the serial Electron Multiplication (EM) register. In practice, spurious electrons from the serial register are indiscernible from and much less frequent than spurious electrons generated in the parallel register. For all practical purposes they can therefore be ignored. Spurious electrons in the parallel register, hereby defined as parallel Clock Induced Charge (pCIC), will be indistinguishable from photoelectrons; they will give rise to a fixed background flux per readout, because the sum of two Poisson distributions is another Poisson distribution. Because of the way the parallel clock signal travels through the parallel register, the distribution of pCIC rates over the image area is particular (Harps\o e, in prep.).

The other place spurious electrons can arise is in the serial EM register. Spurious electrons in the serial EM register, hereby defined as serial Clock Induced Charge (sCIC), are in principle distinguishable from photoelectrons because these electrons will see a shorter length of the EM register, and therefore have lower average gain. Therefore an electron entering the multiplication register at the k'th stage will then logically be described by the random variable $X_k$, where $\gamma_i = (1+p_{\mu})^i$
\begin{equation}
P(X_k = x) = \frac{1}{\gamma_{m-k}}\exp(-x/\gamma_{m-k} )
\end{equation}

\section{Statistical Photon Counting}

It is well known that the number of photons per exposure in one pixel is given by the Poisson distribution, at least if the light is not squeezed. 
{ CCDs are not perfect photon detectors: they only convert photons into photoelectrons with a certain probability, given by the quantum efficiency. The quantum efficiency acts as a perfect neutral density filter, so that the number of photoelectrons is also Poisson distributed and the rate parameter of said distribution is directly proportional to the rate parameter governing the incident photons. Defining $\mathcal{P}$ as the Poissonian Probability Mass Function (PMF) given by
\begin{equation}
\mathcal{P}(n,\lambda) = \frac{\lambda^n e^{-\lambda}}{n!}
\label{poissiondist}
\end{equation}
the probability of $n$ photoelectrons obtained per pixel per readout is given by
\begin{equation}
P(n \mid  \lambda) = \mathcal{P}(n,\lambda)
\end{equation}}
where $\lambda$, not to be confused with wavelength, is proportional to the intensity of the incoming light. One should argue that what we want to measure in order to obtain an image is not the number of individual photons but the intensity of the incoming light; that is, $\lambda$.

{ Henceforth $\lambda$ will refer to the rate parameter of the Poisson process which is the incoming flux of photoelectrons; that is, $\lambda$ is the average number of photoelectrons per pixel per exposure.}

The number of photons per pixel per exposure is not constant for a quiescent source given a constant light intensity, but $\lambda$ is.  
So, if we try to make a better estimate of the light flux level by repeating measurements, we have to take into account the fluctuating number of photons, by estimating $\lambda$ and not the direct number of photons.

\subsection{Probabilistic Analysis of Photons Incident on an EMCCD}

In this section we will fist make an analysis of EMCCD bias frames with pCIC and sCIC electrons. Because of pCIC electrons, a treatment of bias frames will have to include single photon events. As we assume that pCIC electrons are Poisson distributed and the rate of such electrons is sufficiently low that the probability of multiple pCIC electrons in one pixel can be neglected. In later sections we will expand the analysis to account for multiple electron events in the case of photoelectrons. 

In the case that no electron has entered the EM multiplier we assume a constant bias reading; that is, the probability distribution of a bias $B$ is given by
\begin{equation}
P(B=x) = \delta(x)
\end{equation}
{ For simplicity we will assume that the bias is zero, and that bias drift has been corrected by other means, i.e. overscan regions.}
An EMCCD still has conventional additive Gaussian readout noise, but the noise is added after the EM multiplication. Define $R$ as a random variable representing Gaussian readout noise; that is, 
\begin{equation}
P(R=x) = \mathcal{N}(x,\sigma)
\end{equation}
where $\mathcal{N}$ is the normal distribution PDF:
\begin{equation}
\mathcal{N}(n,\sigma) = \frac{1}{\sqrt{2 \pi \sigma^{2} } }e^{-\frac{n^{2}}{2\sigma^{2}}}
\end{equation}
{  We will assume that we obtain a bias reading or the result of an amplified pCIC electron with the probability $p_p \approx \mathcal{P}(1,\lambda_{pCIC})$. Furthermore we will assume that an sCIC electron arises in one of the multiplier steps with probability $p_s$. Because the probability of pCIC and sCIC electrons are very low we will assume that multiple pCIC and sCIC electrons and coincident pCIC and sCIC electrons are negligible. This allows us to write the following, for the total outcome $Z$ of a "bias" reading in the form of a mixture distribution}
\begin{equation}
Z = \left\{ \begin{array}{rl}
 B &\mbox{ with $1-p_p - mp_s$} \\
  X &\mbox{ with $p_p$} \\
  X_k &\mbox{with $p_s$} \\
       \end{array} \right\} + R
\end{equation}
{ Here we will draw a number from $B$ with the probability $1-p_p-mp_s$ or a number from $X$ with probability $p_p$ or a number from one of the $X_{k}$s with probability $p_s$, in either case we will then draw a number from $R$ and add. $X$ is, as previously defined in equation \ref{exponentialphoton}, the stochastic outcome of one EM multiplied electron.
As $B$ and either  $X$ or $X_k$ are mutually exclusive, by definition, and the probability distribution function of a sum of random variables is given by the convolution of the constituting probability distributions,} we can write 
\begin{align}
P(Z=n) & =  \int_{0}^{n} [ (1-p_p-mp_s)\delta(x) \\
             & +  \left(\frac{p_p}{\gamma} e^{-x/\gamma} + \sum_{k=1}^{m} \frac{p_s}{\gamma_{m-k}} e^{-x/\gamma_{m-k}}\right) H(x) ] \nonumber \\
             & \cdot   \mathcal{N}(n-x,\sigma)dx \nonumber
\end{align}
This equation is a mixture distribution of a zero output in the case of no spurious electron and an exponentially distributed output in the case of a spurious electron.  The convolution with $\mathcal{N}$ represents the normally distributed readout noise.

{ In order to use the EMCCD in a photon counting mode the EM gain must be high compared to the readout noise. The convolution on the exponential distributions can therefore be ignored.} We then get 
\begin{align}
P(Z=n) & \approx (1-p_p-mp_s)\mathcal{N}(n,\sigma)  \\
             &+ \left(\frac{p_p}{\gamma} e^{-n/\gamma} +\sum^{m}_{k=1}\frac{p_s}{\gamma_{m-k}} e^{-n/\gamma_{m-k}}\right)H(n) \nonumber
\label{photon}
\end{align}

\begin{figure}
\centering
\includegraphics[width=1.0\columnwidth]{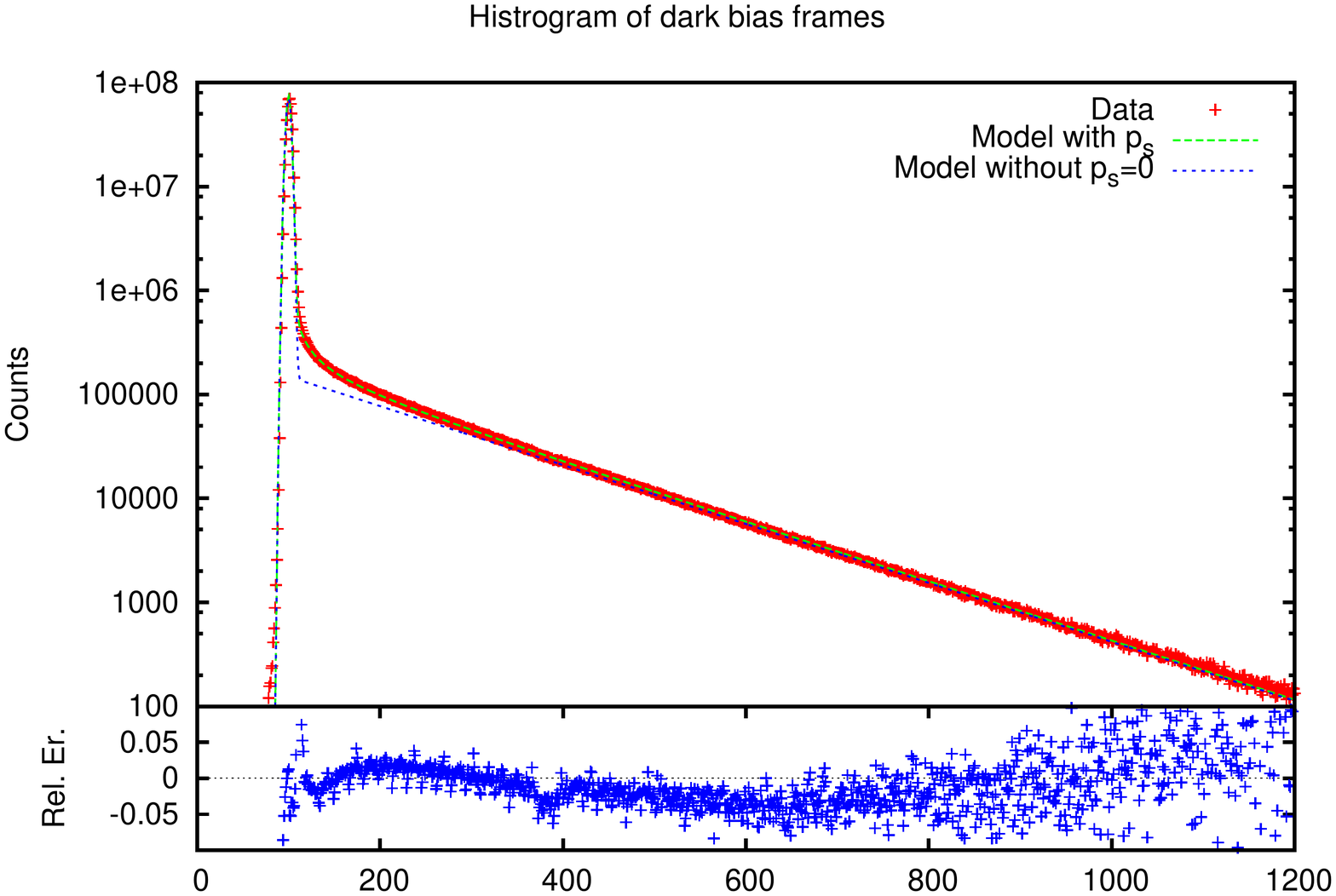}
\caption[Best fit of model to data, with and without the sCIC
term]{{ Histogram of pixel readout values of 500 bias frames, in total
about $5.25 \times 10^{8}$ pixels. The best fit model to the data,
with and without the sCIC term, is over-plotted. The lower panel represent the relative
error between the model with sCIC and the data, defined as
$\delta=(d_i-m_i)/m_i$, where $m$ is the model and $d$ is data.
Because of the large number of pixels in the histogram, even small
effects, which can not be fitted by a simple model, may become
evident. A Fourier analysis of the difference residuals shows a clear
peak of 62 counts at $16$, which is probably from imperfections in the
AD converter. The value from the readout amplifier is a 16bit binary
number, and a period of 16 means that there is an imperfection in the
fourth bit ($2^4 = 16$). In the transition from the bell to the
exponential arm there is a deviation in a small region, which can be
understood as being due to the simplifications done to the equations
to make them computationally manageable.}}
\label{fig:fit}
\end{figure}

In Fig. \ref{fig:fit} the model in equation \ref{photon} has been fitted to a histogram of a series of bias frames from an EMCCD, { as the green curve. The blue curve is a fit where $p_s$ was forced to zero to highlight the effect of sCIC.}

\subsection{Thresholding}

If the flux is sufficiently low, then the probability of having a photo-electron in a pixel is small, implying that most pixels will contain zero electrons. Note that in fact we can always configure the instrument so that the former is true. If there are no electrons in a pixel, there is nothing to cause an electron avalanche in the multiplier apart from sCIC electrons. { This implies that we can choose a threshold $T$ so that if the readout from a pixel is below this threshold there is probably no photo-electron in the pixel. And if the readout is above the threshold, there is probably one or more electrons in the pixel. Due to the statistical nature of the multiplication in an EMCCD, it might not be possible to count every individual photo-electron, but we can make an accurate estimation of the rate of photoelectrons, $n$, in a single pixel over a series of exposures with fixed integration time.}

One way to estimate $\lambda$, via thresholding, is by dead time correction. In doing dead time correction we use the fact that if we detect a pixel value over a certain wisely chosen threshold $T$, we know that, with a high probability, there are photons in that pixel, but we do not know how many, due to the stochastic nature of the cascade amplification. { On the contrary, if we see a pixel value below the threshold we know, with a high probability, that there are no photoelectrons in that pixel and obviously zero photon-electrons is a uniquely defined case, as opposed to the former, which could contain one or more photon-electrons.} However, irrespectively of how the threshold is set, there will always be a finite probability that a pixel above the threshold contains no photon-electrons, and visa versa, that a pixel below the threshold contains photon-electrons.

Thus, in the idealised case, if we observe $N$ frames and, for a certain pixel, see $Q$ frames with photons in that pixel, then we can estimate the intensity $\lambda$ from equation \ref{poissiondist} and the frequency of non detections
\begin{align}
\frac{N-Q}{N} & =  \mathcal{P}(0,\lambda ) &  \\
               & =  e^{-\lambda} &\Rightarrow  \nonumber \\ 
      \hat{\lambda} & =  -\operatorname{ln}\left( \frac{N-Q}{N} \right) &
      \label{eq:deadtime}
\end{align}
Note that because of the pCIC electrons, this estimated $\hat{\lambda}$ will be the sum of the pCIC rate and the photoelectron rate, which is the flux.

The readout noise $R$ could rise above the threshold and give false counts. The probability of $R$ rising above the threshold can be calculated as 
\begin{equation}
P(R>T) = \int^{\infty}_{T} \mathcal{N}(n,\sigma) dn = \frac{1}{2}-\frac{1}{2}\operatorname{erf} \left( \frac{T}{\sigma \sqrt{2}} \right)
\end{equation}
A pixel will be above the threshold with the probability $(1/2)(1-\operatorname{erf} (T/\sigma \sqrt{2})$, just due to readout noise. Expressing in terms of $\hat{\lambda}$ we get 
\begin{equation}
\hat{\lambda} = -\operatorname{ln}\left(\frac{N-\left[ Q + \frac{N}{2}\left(1-\operatorname{erf}  \left( \frac{T}{\sigma \sqrt{2}} \right)\right)\right]}{N}\right)
\end{equation}
The last term under the logarithm is small, so we can expand the logarithm as $ \operatorname{ln}(a+x) \approx \operatorname{ln}(a) + x/a$.
\begin{equation}
\hat{\lambda} = -\operatorname{ln}\left(\frac{N-Q}{N}\right) + \frac{N \left( 1-\operatorname{erf}  \left( \frac{T}{\sigma \sqrt{2}} \right) \right)}{2(N-Q)}
\end{equation}
We will get a contribution of $\frac{N \left( 1-\operatorname{erf}\left( \frac{T}{\sigma \sqrt{2}} \right) \right)}{2(N-Q)}$ to $\hat{\lambda}$.

{ There is a finite probability that a single photo-electron does not get amplified above the threshold, given by
\begin{equation}
P(X<T) = 1-e^{-T/ \gamma }
\end{equation}
which will cause a bias in the estimator.}

This thresholding method is sufficiently simple that we can actually analytically calculate the mean and variance of the estimation on $\lambda = \lambda_{p}+\lambda_{pCIC}$, where $\lambda_p$ is the rate of photoelectrons and $\lambda_{pCIC}$ is the rate of pCIC electrons per readout, thus rendering it possible to calculate bias in the estimator. For a Poisson distribution the probability of getting anything above zero is $1-e^{-\lambda}$. If we neglect multiple electron events, we can assume that an electron will be amplified above the threshold with the probability $e^{T/ \gamma}$. If we record N frames, then the number of pixels we will get above the threshold will be given by the binomial distribution, with N trials and the probability of ``success'' given by
\begin{align}
p&=e^{-T/ \gamma }(1-e^{-\lambda}) \label{psigma} \\
 & +\frac{e^{-\lambda}-mp_{s}}{2}\left(1-\operatorname{erf}  \left( \frac{T}{\sigma \sqrt{2}} \right) \right) \nonumber \\
 &+ p_s\sum_{k=1}^{m} (1-e^{-T/\gamma_{m-k}})  e^{-\lambda} \nonumber
\end{align}
To obtain a manageable expression we have here disregarded the possibility of coincident pCIC and sCIC events, as a few extra electrons in the EM multiplier steps when a photoelectron is already generating an electron cascade is insignificant.
To get the bias and variance on the estimated $\lambda$, we can simply calculate the mean and variance of our estimate $\hat{\lambda}$ over the appropriate binomial distribution as done below.
\begin{equation}
E(\hat{\lambda}) = -\sum_{Q=0}^{N-1} \operatorname{ln}\left(\frac{N-Q}{N}\right)  \binom{N}{Q} p^{Q} (1-p)^{N-Q}
\label{eq:thresholdmean}
\end{equation}
\begin{equation}
Var(\hat{\lambda}) = \sum_{Q=0}^{N-1} \left( -\operatorname{ln}\left(\frac{N-Q}{N}\right) - \hat{\lambda} \right)^{2} \binom{N}{Q}  p^{Q} (1-p)^{N-Q}
\label{eq:thresholdvariance}
\end{equation}
It should be noted that the lower limit for detectable flux with this method is not given by the readout noise, even though it enters in the form of $\sigma$ in equation \ref{psigma}. Instead the lower limit is given by $\lambda_{pCIC}$, because the pCIC events are indistinguishable from photoelectrons. The pCIC events will generate shot noise, which will dominate lower signal fluxes.

Theoretically the estimator bias is a nonlinear function of $\lambda$, due to multiple electron events. The probability of surpassing the threshold is significantly larger for multiplication cascades generated by multiple electrons than for cascades generated by a single electron, thus introducing a nonlinear bias. It is possible to correct for this effect numerically, as equation \ref{psigma} can be modified to include multiple events via the Cumulative Distribution Function (CDF) for the Erlang distribution: 
\begin{align}
p & = \sum_{k}\sum^{k-1}_{n=0} \frac{(\gamma T)^{n})\lambda^{k}e^{-\gamma T -\lambda}}{n!k!}   \\
   &+\frac{e^{-\lambda} - mp_{s}}{2}\left(1-\operatorname{erf}  \left( \frac{T}{\sigma \sqrt{2}} \right)\right)e^{-\lambda} \nonumber \\
  &+ p_s\sum_{k=1}^{m} (1-e^{-T/\gamma_{m-k}})  e^{-\lambda} \nonumber
\end{align}

Fortunately this effect is small, as the number of multiple events in the range of $\lambda$, where this method is applicable, is very small.

\subsection{Full Bayesian Inference}
One particular problem with thresholding is that, in principle, we lose information in the process: because the strength of the signal before thresholding actually contains information, a stronger signal will be correlated with a higher number of photons, albeit not directly proportional to a higher number. The correlation between the readout signal and the number of photons is given by Erlang distributions as in fig. \ref{fig:ProbPlot}.
Combining the desire to estimate $\lambda$, not the number of photons, with the desire to avoid thresholding naturally leads to Bayesian inference. With this method we can combine all of the probability distributions involved in the data recording process, and through this estimate $\lambda$ without thresholding.

Equation \ref{photon} outlines a compelling way of counting photons in a series of EMCCD frames; that is, attain the real shot noise limit, with Bayesian methods. 

For photons incident on a pixel at some rate $\lambda$, the amount of photons (and hence photoelectrons) accumulated in the pixel in a given time frame is given by equation \ref{poissiondist}. With this information we can expand the distribution of amplified electrons in equation \ref{photon} to read
\begin{align}
P(n \vert \lambda) & = ((1-mp_{s})\mathcal{N}(n,\sigma) \label{photon21} \\
 &+ p_{s}\sum_{k=1}^{m}\mathcal{G}(n,0,\gamma_{m-k}))\mathcal{P}(0,\lambda ) \nonumber \\
 &+ \sum_{k=1}^{\infty} \mathcal{G}(n,k,\gamma) \mathcal{P}(k,\lambda) \nonumber
\end{align} 
$\mathcal{G}$ is a gamma distribution given by
\begin{equation}
\mathcal{G}(n,k,\gamma) = n^{k-1} \frac{e^{ -\frac{x}{\gamma} } }{\gamma^{k} \Gamma (k) }
\end{equation}
This is simply a mixture distribution over the probabilistic results of reading out zero, one, two,... photoelectrons, where the mixture coefficients (which is the probability of drawing from the corresponding distribution) is given by an appropriate Poisson distribution. { Furthermore in the case of zero photoelectrons we will draw either a sCIC electron from one of the EM amplifier steps, or read noise. This equation does not take into account readout noise on electron events, which is the same approximation as in eq. \ref{photon}, or the possibility of coincident photo- and sCIC electrons.
In the general case the terms in eq. \ref{photon21} should be convolved as the events of photo- pCIC and sCIC electrons can occur together and they add, but the convolution of gamma distributions with differing scale parameters do not have any closed form expression. There are approximations in terms of other gamma distributions though see  for instance \cite{Stewart2007}. In this case because the EM gain is large compared to the readout noise and because sCIC electrons do not contribute significantly to the electron cascades already being generated by pCIC- or photoelectrons, this approximation as a mixture distribution is appropriate. In the appendix, these distributions are expressed in terms of phase type distributions which can, in principle, express the convolved distribution  analytically  in terms of exponentials of matrices.}

{ We will assume that pCIC electrons will be Poisson distributed as the event of a pCIC electron is rare and integer.} Therefore pCIC electrons can formally be included in $\mathcal{P}$, as the sum of two independent Poisson stochastic variables is Poisson distributed with a rate parameter that is the sum of the two constituent rate parameters. 

Given a series of measurements of $n$, $\lambda$ is the number we want, so we will inverse the given distribution via Bayes formula.
\begin{equation}
P(\lambda \vert n) = \frac{P(n \vert \lambda)P(\lambda)}{\int P(n \vert \lambda )P(\lambda)d \lambda}
\label{bayes}
\end{equation}
The denominator in this integral looks discouragingly complicated but can be integrated term by term, and for any reasonable flux level, only the first 10 terms or so will be significant.  

Assuming a constant rate of photons (photoelectrons), the rate can be estimated from a series of electron counts $\{ n_{i} \} $ as 
\begin{equation}
P(\lambda \vert \{ n_i \} ) = \prod_i P(\lambda \vert n_i )
\end{equation}

One thing we hitherto skipped is the prior probability distribution $P(\lambda)$, which would express how much we believe in a particular value of $\lambda$ before any data becomes available to us. In statistical literature there is often extensive discussions on how to choose a good prior \citep{Gregory2005}. Our choice of prior depends on what we believe about $\lambda$ in the first place. If we believe that every value of $\lambda$ is equally probable, we could use the Laplace principle of indifference and assign a prior probability of $1/n$ to each of the $n$ $\lambda$s. This assumes that there are only finitely many possible values $\lambda$, which might be a good assumption given that we ultimately discretise the problems to do the calculations on a computer. 
We could also believe that the prior probability should be equally distributed on all scales; that is, the units we use on $\lambda$ should not influence our conclusions. This is also a very reasonable assumption and leads to the whole concept of Jeffrey's prior \citep{Gregory2005}. 

Fortunately, we don't have to care much about the prior in many experiments, and neither do we need to in this experiment. A prior must almost per definition be a slowly variable function; a peaked prior would indicate that we already know a lot about the parameter we try to estimate and there would not be any reason to actually do the experiment. This means that the prior only becomes important if the likelihood function is not very peaked, which in turn would mean that the individual data points do not contribute with very much information. That is, the experiment is badly designed for measuring the parameter we are interested in, { or if only very few data points are available. This is not the case here: the likelihood function is peaked and we presume that a significant amount of data points are available, thus the prior becomes unimportant.}

For high framerate applications like lucky imaging, assuming a constant rate might not be appropriate; fortunately, the total amount of accumulated photons over a series is simply the sum of the rates. It is always true that the distribution of a sum of independent random variables is the convolution of their individual distributions. So regarding each $P(\lambda \vert n_i )$ as the probability distribution for the estimator $\lambda_i$ of the rate in each individual sample, we can construct the probability distribution of $I = \sum_i \lambda_i$ as 
\begin{equation}
P(I \vert \{ n_i \} ) = \bigotimes_i P(\lambda \vert n_i)
\label{bayes2}
\end{equation}
where $\otimes$ is taken to mean the convolution of all the individual distributions. 

The independence of the $P(\lambda \vert n_i )$s is insured by the memorylessness property of Poisson processes, stating that the number of events in any number of non-overlapping sections of a Poisson process will be independent. This independence also implies that there can not be any requirement for the $P(\lambda \vert n_i )$s to be consecutive in time it is perfectly possible to choose which samples to include, in correspondence with the principles of lucky imaging.

If we are only interested in simple Maximum A Priori (MAP) estimates of the rates, we don't have to calculate the denominator in equation \ref{bayes} and \ref{bayes2}, as both the product operator and the convolution operator are linear. The denominator normalises the expression but is not needed for a MAP estimate.

%% file: Results.tex
\section{Test Results}

The CCD used for the experiments in this article is an EMCCD in an Andor iXon$^{EM}$+ 888 camera, which employs an E2V CCD201 with 1024 by 1024 $13 \times 13 \mu m$ pixels; it is an electron multiplying frame transfer CCD. We have used a readout speed of $10MHz$, with a corresponding readout noise of 50 to 100 electrons, but because of the electron multiplication gain this relatively high readout noise will not be a problem.
In the EMCCD used, the EM register has 604 stages. The average amplification will then be on the order of $\gamma = (1+p_{m})^{604} \approx 500$, if $p_m \approx 0.01$. If one photo-electron on average generates a signal of $\approx 500$ electrons, the signal from one photo-electron will in most cases be easily recognisable, even if the readout amplifier has a readout noise of about 60 electrons, which is the case here. The characteristics of the camera, i.e. EM gain and bias level, seem to fluctuate, presumably largely due to temperature changes when the electronics heat up during long series of fast readouts.

The iXon camera was chosen for the experiment because it is commercially available. The rate of pCIC, which is the dominant detector-induced noise source in the final data, is on the order of one per 20 pixels read out in this particular camera. This is fairly high compared to state-of-the-art research EMCCD controllers, which can have pCIC rates down to 0.1\% see for instance \cite{Daigle2010}. The high pCIC rate of the iXon camera is not a problem for the tests reported here. 

It is difficult to make a calibrated light source that emits about one photon per second per pixel. But because of the switching time of an Light Emitting Diode (LED), sufficiently short bursts of light can be made. By flashing a LED a certain number of times per exposure, a sufficiently low and repeatable flux level can be achieved.
By this approach, the number of photons captured will be proportional to the number of flashes, or rather, the rate parameter $\lambda$ will be proportional to the number of flashes. For our setup, flashes with a duration of $2ms$ from a LED placed $2m$ from the chip produced appropriate flux levels.

\subsection{Linearity}

\begin{figure}
\centering
\includegraphics[width=1.0\columnwidth]{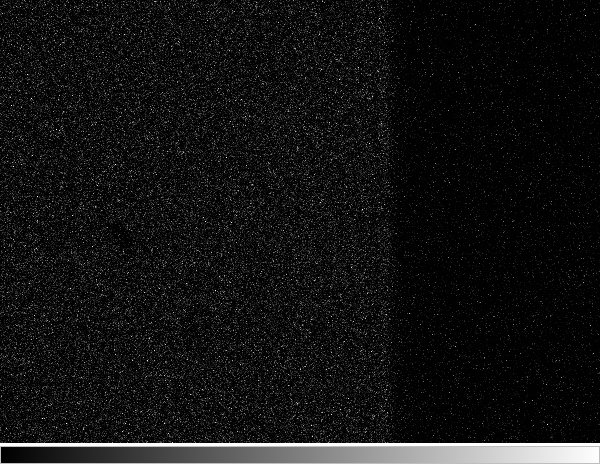}
\caption[Image from EMCCD]{Image captured by the EMCCD when it was exposed for 1 second with 10 flashes and a multiplier gain of 1000. The right third of the image was covered with black tape to provide a bias reference.}
\label{fig:LambdaImage}
\end{figure}

\begin{figure}
\centering
\includegraphics[angle=0,width=1.0\columnwidth]{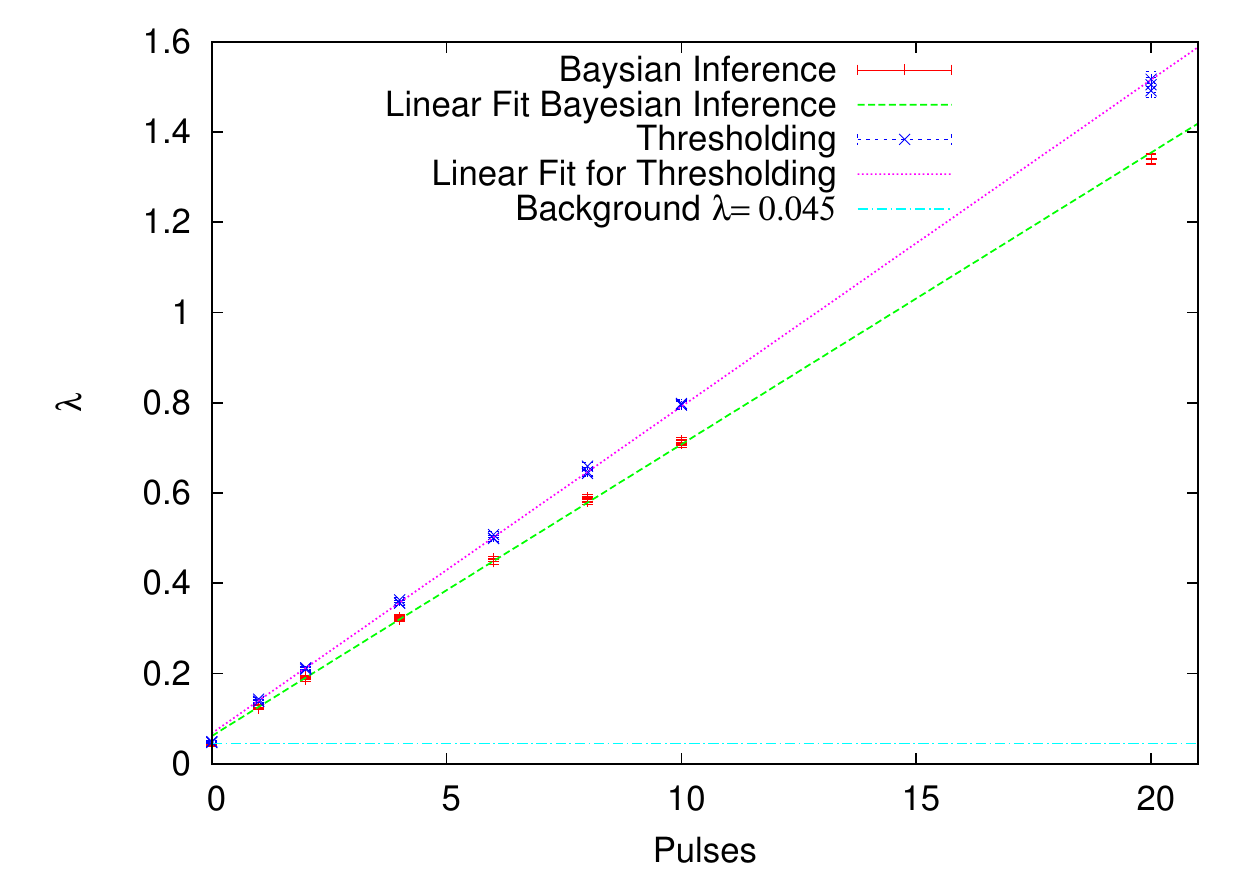}
\caption[Test of linearity]{Test of the linearity of the maximum likelihood and thresholding methods: they both behave very linearly, as expected, but with different background levels. The thresholding method correction reports a higher background level and it breaks down for $\lambda$ higher than approximately $1.5$, whereas the MaxLik method is linear up until approximately $\lambda=10$ (not shown on the graph). The uncertainty on $\lambda$ rises with $\lambda$ as expected. The Bayesian Inference method seems to slightly underestimate $\lambda$. This method contains more parameters than the thresholding method, one of them being the EM gain. A slight overestimation of the EM gain would explain the underestimation of $\lambda$.}
\label{fig:LinearPulsPlot}
\end{figure}

To test the methods for reducing data from EMCCDs, the chip was uniformly illuminated with a green flashing LED, with half of the chip covered with black tape, as shown in Fig. \ref{fig:LambdaImage}. The timing and the number of flashes was controlled by a custom-built delay timer that would also control the triggering of the camera via the camera's trick input terminal. 
A series of 100 images was captured for each rate parameter, i.e. number of flashes. The dark half of the image is used for estimating the bias via a median filter on a line-by-line basis, as this particular type of camera can display rather large fluctuations in the bias level.
For evaluating the thresholding and Bayesian inference reduction methods, 150 pixels in a single row were selected, { as bias variations can be eliminated, by estimating the bias from the dark part of said line.} To test the Baysian inference with the maximum likelihood MAP estimator, a series of dark frames was also captured to adjust all of the parameters in the model for the dark response of the camera. The results for this method are shown in Fig. \ref{fig:LinearPulsPlot}.

The background level in Fig. \ref{fig:LinearPulsPlot} has been established by applying the thresholding and the Bayesian inference (MaxLik) algorithm to a set of dark frames. Both algorithms reported $\lambda \approx 0.045$. This background floor is related to the rates of pCIC electrons and sCIC electrons (hence they have more or less the same numerical value). This contribution is constant per readout; it is, fully equivalent with an extra background flux, thus adding noise. { If the probability of pCIC and sCIC events is lowered, then the detector-induced stochastic background is also lowered.}
The thresholding algorithm applied to the same data gives very similar results until $\lambda$ reaches 0.5; that is, for low levels of light the results are linear as seen in Fig. \ref{fig:LinearPulsPlot}. { Only data up to 20 flashes are included, as the thresholding algorithm breaks down when applied to the data set with 50 flashes and reports infinite values for $\lambda$.}

\subsection{Signal to Noise Ratio}

\begin{figure}
\centering
\includegraphics[angle=-90,width=1.0\columnwidth]{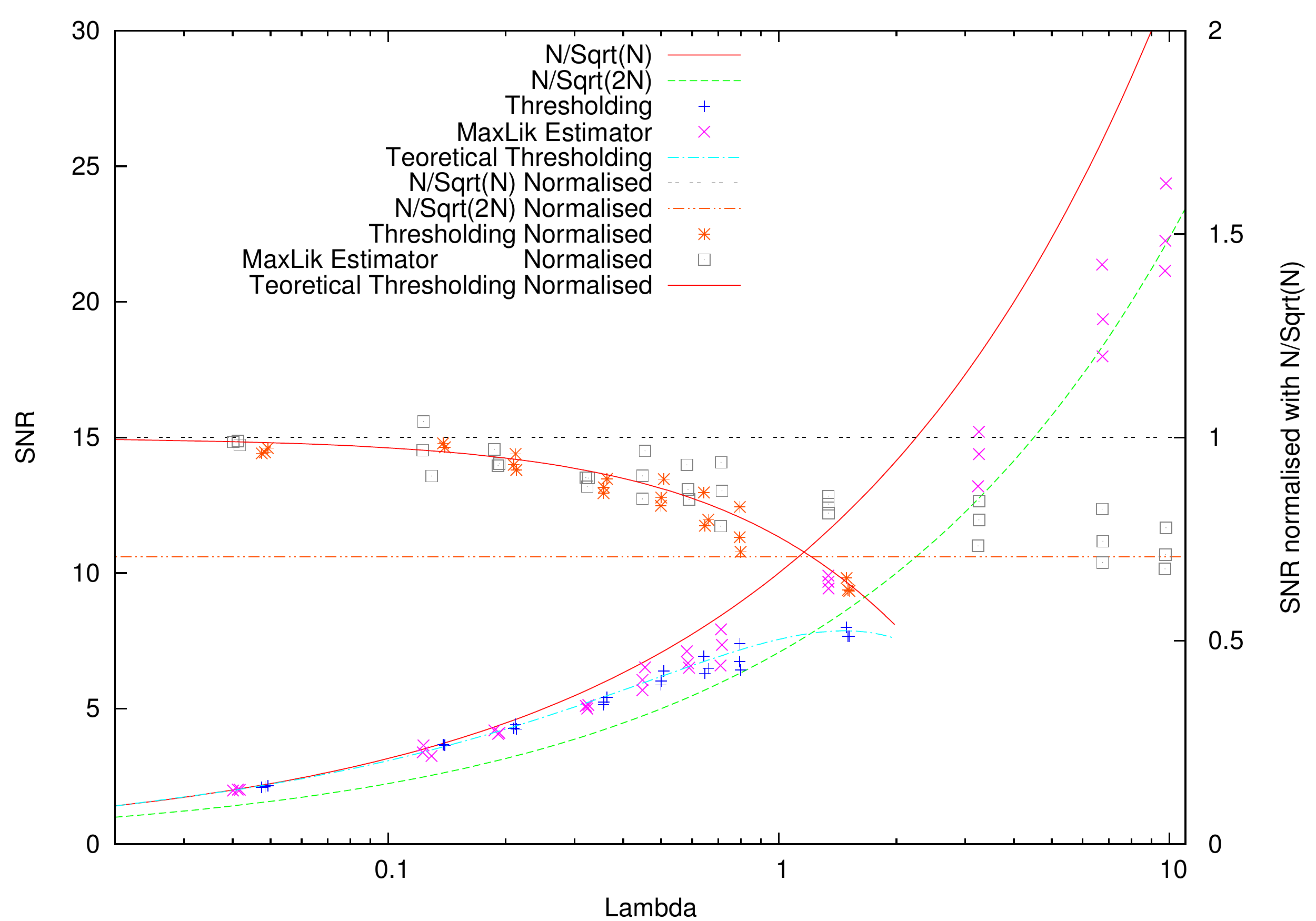}
\caption[Test of the S/N for rising $\lambda$]{Test of the S/N for rising $\lambda$. The two lines represent the theoretical shot noise limit and the theoretical asymptotic $\sqrt{2}$ excess noise of an EMCCD. It seems that the two methods are comparable up until about $\lambda = 1$ where the performance of the dead time correction deteriorates as expected. The theoretical curve has been calculated based on the equations \ref{eq:thresholdmean} and \ref{eq:thresholdvariance}. The data set with the lowest $\lambda$ has been obtained from a set of dark frames. The electrons in this set are mostly parallel CIC electrons but they follow the same statistics and are completely indistinguishable from photo-electrons.}
\label{fig:BayesDeathTimePlot}
\end{figure}

The signal to noise ratio can be estimated from the sample variance and mean of $\lambda$ as:
\begin{equation} 
S/N = \frac{\mu(\lambda)}{\sigma(\lambda)}
\end{equation}
This was done for all of the data sets; see Fig. \ref{fig:BayesDeathTimePlot}. For $\lambda$s less than about 0.5, both methods achieve an excess noise close to unity. There is good agreement between the theoretical model of the thresholding method and the thresholded data. The thresholding method is just as good at estimating $\lambda$ up to about one, where the thresholding method deteriorates as expected, due to coincidence loss. 
There is some noise in the estimates of the S/N ratios. The Bayesian inference method in particular is very sensitive to its parameters, especially the bias level. Unfortunately the bias level in this particular camera fluctuates, which could explain the spread in the calculated S/N ratios.

\section{Discussion}

We have presented two methods for analysis of EMCCD data. We have shown that these two methods are linear, and that at fluxes in the range $\lambda=0.05$ to $\lambda \approx 1$, they achieve the shot noise limit (see Fig. \ref{fig:BayesDeathTimePlot}).  The noise rises above this limit to reach the ordinary excess noise limit of two at $\lambda \approx 3$. 

If we define photon counting as achieving the shot noise limit, then we have shown in the analysis of our EMCCD data that it is in fact possible to do photon counting with an EMCCD, under certain circumstances. Both the thresholding and the Bayesian maximum likelihood method can count photons when the light flux is sufficiently low: less than about half of a photon per pixel per readout. Our definition of photon counting is statistical. We cannot point to and count the individual photons because of coincidence loss, but we can count them in terms of rates ($\lambda$s), in the sense that over a series of exposures we can accurately, down to the fundamental shot noise limit, estimate rates much less than one photon per exposure. It is important to note this distinction. Every time we try to estimate the number of photons over a number of exposures, we do estimate rates in reality; making the statistics reflect this makes the derivations much more clear. This is why the theory has been developed in terms of rates. 

The thresholding method is slightly more robust to fluctuating bias levels in the data, but fails by design when the flux reaches a few photons per pixel per readout, and at $\lambda$ above one it is no longer optimal. The Bayesian inference method, on the other hand, also counts photons, but does not imply thresholding-- therefore it does not break down at high fluxes. But it does have excess noise at these fluxes, and at fluxes reaching about three photons per pixel per readout there is no benefit from photon counting compared to just simple integration with an EMCCD.

In general the proposed methods are not affected by readout noise; instead, the presence of spurious CIC electrons at the level of 0.045 per pixel read gives a significant contribution to the background, given that $\lambda$ preferably should be below 1. This problem can effectively be completely avoided by using readout electronics that have been optimised for low CIC rates.

One thing to note is that in the Bayesian interpretation of probability, probability simply expresses degree of certainty. It makes explicit the knowledge that is taken into account to produce an output degree of certainty. So the reason that we here can lift some of the degeneracies from the gamma distributed output of the EMCCD is that we explicitly add the knowledge that photons are Poisson distributed. Conversely we can now be certain that there is no more information about the flux in the data, unless we add some more external information about the EMCCD or photons.

The two methods proposed have different strengths and weaknesses. The Bayesian inference method is more sophisticated and elegant, it does not inherently break down at a certain flux level, and it has shown slightly better results in the lab. However, it is computationally heavy, and sensitive to parameter estimates. The thresholding method is much lighter computationally (orders of magnitude faster), and it is accurate at low light levels.
The methods proposed are, with computer science terms, embarrassingly parallel. This refers to the fact that there is no interdependency between the pixels in the frames with respect to counting photons. It should therefore be fairly simple to implement these algorithms on a number of parallel processors. We simply assign one pixel to each processor. The individual calculations are also fairly simple and it might be possible to implement the algorithms on some kind of stream processor. Stream processors are a simplified kind of processor often found in GPUs (graphics processing units). One GPU can contain many stream processors. In essence it could be possible to processes the photon counting algorithms on a high-end graphics board and have many parallel processors available.

Both of the methods used are statistical in nature and it is therefore necessary to obtain a large number of frames for every object that one wants to study, implying that large amounts of data needs to be recorded. A thresholding method might be an advantage in this respect. We could store only the thresholded frames; they will contain only ones and zeros (mostly zeros) and could therefore be compressed by one to two orders of magnitude.

Other ways to compress the data could be to still do thresholding and substitute everything below the threshold with zero, but keep the value in a pixel if the value is above the threshold. This would ensure that the majority of the pixels containing only readout noise would be zero, but we would keep the information above the threshold. Because most of the pixels would then have the same value (zero), we could compress the data significantly. It is important to note that we could make even the Bayesian inference method work on these data if we modified the probability model behind it. This scheme could also be used to implement a hybrid thresholding/simple integration algorithm, where thresholding is used in pixels with low flux and simple integration in pixels with high flux, as simple integration actually introduces less noise than the thresholding method at high fluxes. 

The application of EMCCDs for photon counting is quite specialised in the sense that the detector has to be severely photon-starved before an EMCCD has the upper hand compared to modern low noise CCDs. One area where photon counting with EMCCDs might be applicable is in high frame rate imaging such as lucky imaging. 

The readout noise in a CCD tends to rise with the readout speed. The readout noise in an EMCCD will rise with the readout speed as well, but because of the multiplication, the effect of readout noise is negligible. Another characteristic of high frame rate imaging is that the number of photons in each pixel per readout tends to be low, simply because of the low exposure time. In the methods we have developed here, there are no constraints on which frames out of a series of frames one includes in the estimation of $\lambda$. This means that one probably could use these photon counting algorithms in combination with the principles of lucky imaging. 

If one can obtain independent information on which frames to include due to instantaneous good seeing, one could pick these frames and run one of the photon counting algorithms in these selected frames and estimate $\lambda$. In principle, this would enable one to make much more sensitive lucky imaging exposures, as one could detect $\lambda$s less than one photon per frame. The independent information on the seeing could come from a bright star in the frame. If the star is bright enough, it will be possible to see the seeing effects on this star without photon counting. This will be at the cost of a factor of two in excess noise, but if the star is bright enough, this is not important. In a region around this star, the seeing will be close to the seeing affecting the bright star, and we could do lucky imaging combined with photon counting in this region.

%% file: EMteori.tex
\section{Probability Distribution of EMCCD Output with Serial Clock Induced Charge}

To describe the distribution of electrons coming out of a typical electron multiplication register found in EMCCD or L3CCD, we will here use the theory of phase type probability distributions. For a thorough discussion of these distributions see \cite{Latouche1999}. 

\begin{figure}
\centering
\includegraphics[width=1.0\columnwidth]{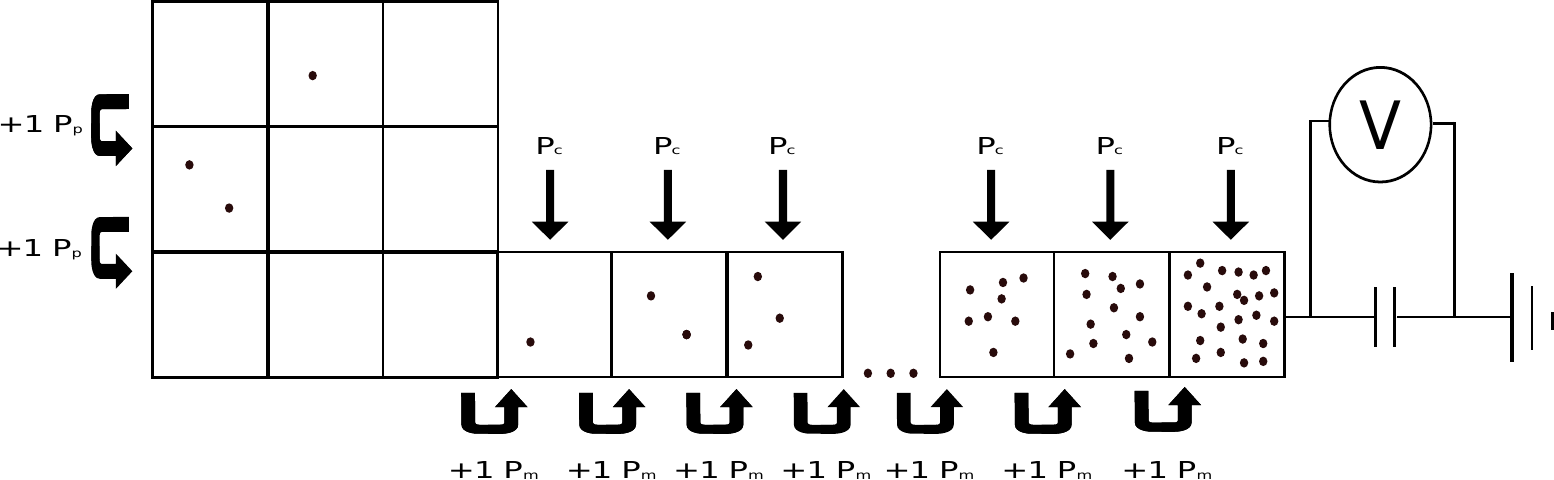}
\caption[Drawing of noise sources]{Principal drawing of the different sources of noise and amplification in an EMCCD. In the serial register we add an electron to every electron we shift though the pixels with the probability $p_{\mu}$; this is what makes the amplification. Also, for every pixel in the serial register, we add one electron with the probability $p_s$. This is called serial Clock Induced Charge (sCIC). In the parallel register, we also add an electron to every pixel per shift with the probability $p_p$. This is called parallel Clock Induced Charge (pCIC), which unfortunately will be indistinguishable from dark and photo electrons, except that dark and photo-electrons will be proportional to exposure time and pCIC will not.}
\label{fig:EMCCDprob}
\end{figure}

If we have a register where there is a probability $p_{\mu}$ that the one electron is multiplied in one stage of the multiplier, then the multiplication gain of $M$ steps is given by
\begin{equation}
\gamma_m = (1 + p_{\mu})^m
\end{equation}
The multiplication of one input electron can then be described by the random variable $X_M$, where 
\begin{equation}
P(X_m = x) = \frac{1}{\gamma_m}\exp \left(-\frac{x}{\gamma_m} \right)
\end{equation}
So an electron entering the multiplication register at the k'th stage will then logically be described by the random variable $X_i$, where $i=m-k$
\begin{equation}
P(X_i = x) = \frac{1}{\gamma_i}\exp \left(-\frac{x}{\gamma_i} \right)
\end{equation}
We can use these equations to model another source of noise in the electron multiplier, namely serial Clock Induced Charge events (sCIC). We will model this source of noise as an electron entering the k'th stage with the probability $p_s$, see Fig. \ref{fig:EMCCDprob}. If $p_s$ is sufficiently small, the probability of two or more sCIC events in the multiplier during the multiplication of one electron is negligible. 
This effect will give a non trivial-output distribution even for a null electron input. 

We will exclude the possibility of more than one sCIC electron in each step so the probability of getting $Y=y$ from a particular step $i$ must be given by $p_s P(X_i = y)$. We also exclude the possibility of getting a sCIC electron in more than one step. This means that we can simply add the probabilities of $Y=y$ for each step based on the rule that we can add probabilities for mutually excluding events.
That is, we see $Y$ as a number that we draw from the distributions $X_i$ according to this table:
\begin{center}
\begin{tabular}{|c|c|}
$Y$ & $p$ \\
\hline 
$0$ & $1-m p_s$ \\
$X_{m-1}$ & $p_s$ \\
\ldots & \ldots \\
$X_1$ & $p_s$ \\
\end{tabular}
 \end{center}
This way of choosing between different distributions with different probabilities is known as a density mixture.
Summing up we get the following distribution for $Y$
\begin{equation}
P(Y=y)=
\begin{cases}
(1-mp_s) , \, y=0\\
p_s \sum_{i=1}^{m}P(x_i=y) , \, y>0
\end{cases}
\end{equation}
This kind of distribution for $y>0$ is known in the mathematical literature as a hyper-exponential distribution, or a density mixture of exponential distributions. It is efficiently described within the framework of phase type distributions \citep{Latouche1999}. 
{ The PDF of such distribution is characterised by its so-called $\hat{S}$ matrix.}
If we define $g_i = 1/\gamma_i$, then we can write 
\begin{equation} 
\hat{S} = \left( \begin{array}{cccc}
  -g_{1} & 0 & \cdots & 0 \\
  0 & -g_{2} & \cdots & 0 \\
  \vdots & \vdots & \ddots  & 0 \\
  0 & 0 & \cdots  & -g_{M}\\
  \end{array} \right) 
\end{equation}
The probability vector for this distribution is then given by 
\begin{equation}
\bar{\alpha} = p_s \bar{1}
\end{equation}
As we can see, $\bar{\alpha}$ is a vector of the probabilities in the previous tables, and the $g$'s correspond to the different exponential distributions of the $X_i$'s, as the only difference between them is their gain $\gamma_i$.
From the theory of phase type probability, we then have that
\begin{equation} 
P \left(Y = y\right) = \bar{\alpha} \exp(y \hat{S}) \bar{S_0}
\end{equation}
where $\bar{S_{0}} = -\hat{S} \bar{1}$

The factor $\exp(x\hat{S})$ is the matrix exponential which is defined in terms of its series expansion, so 
\begin{equation}
\exp(x\hat{S}) = \sum_{i=0}^{\infty} \frac{(x\hat{S})^i}{i!}
\end{equation}
Taking a product of a matrix with itself is a well-defined operation, so this exponential is also well-defined. As $\hat{S}$ is diagonal, we can write this out and we get 
\begin{equation} 
\exp(x\hat{S}) = \left( \begin{array}{cccc}
  \exp(-xg_{1}) & 0 & \cdots & 0 \\
  0 & \exp(-xg_{2}) & \cdots & 0 \\
  \vdots & \vdots & \ddots  & 0 \\
  0 & 0 & \cdots  & \exp(-xg_{M})\\
  \end{array} \right) 
\end{equation}
We then get 
\begin{equation}
\bar{\alpha} \exp(y \hat{S}) \bar{S_0}  = p_{c}\sum_{i}^{M} \frac{1}{\gamma_i}\exp \left(-\frac{y}{\gamma_i} \right)
\end{equation}
Which is exactly the distribution for $P(Y=y)$ when $y > 0$.
The term describing the zero case where we obtain no sCIC electrons, $y=0$, can also be included directly in the distribution by including a $g_{0}$, and taking the limit $g_{0} \rightarrow \infty$ as
\begin{equation} 
\lim_{g \rightarrow \infty} g \exp(-g x) H(x) = \delta(x)
\end{equation}
where $H(x)$ is the Heaviside function. The Heaviside function takes care of the negative side of the exponential function, as there is no such thing as negative probability. By including this limit and changing the first entry in $\bar{\alpha}$ to $1-m p_s$, we obtain the full probability distribution for $Y=y$.

It is also interesting to note that we can calculate the total distribution with one electron going through the multiplication register with CIC noise included within the framework of phase type distributions. We will define random variable $Z$ to be the total number of output electrons.
\begin{equation}
Z = X_M + Y
\end{equation}

We can think of this number as being drawn from different distributions according to this table:
\begin{center}
\begin{tabular}{|c|c|}
$Z$ & $p$ \\
\hline 
$X_m$ & $1-m p_s$ \\
$X_m + X_{m-1}$ & $p_s$ \\
$\ldots$ & $\ldots$ \\
$X_m + X_1$ & $p_s$ \\
\end{tabular}
\end{center}

The first line is the case in which we obtain no sCIC electrons in a run through all the stages. Line number two is the case in which we get a sCIC electron in the stage directly proceeding the first stage and so on.
The distribution of a sum of random variables that are exponentially distributed is known in the mathematical literature as a hypo-exponential distribution, so the distribution of $Z$ is a density mixture of exponential and hypo-exponential distributions. This can also be conveniently plugged into an $\hat{S}$ matrix in the following way
\begin{equation}
\hat{S} = \left( \begin{array}{ccccccc}
  -g_{M} & 0            & \cdots               & 0                          & \cdots              & 0            & 0     \\
  0            & -g_{M} & g_{M}   & 0                         & \cdots               & 0            & 0     \\
  0            & 0            & -g_{M-1}         & g_{M-1} & \cdots               & 0            & 0     \\
  \vdots       & \vdots       &  \vdots                   & \vdots                 & \ddots              & 0            & 0     \\
  0            & 0            &  0            & 0                & \cdots                 & -g_{M} & g_{M}  \\
  0            & 0            &  0            & 0                & \cdots                 & 0                       & -g_{1} \\
  \end{array} \right) 
\end{equation}
The $\bar{\alpha}$ vector corresponding to this $\hat{S}$ matrix is given by
\begin{equation} 
\bar{\alpha} = [ 1-mp_s, p_s, 0, \cdots, p_s, 0 ]
\end{equation}
The distribution of $Z$ will then as before be
\begin{equation} 
P \left(Z = z\right) = \bar{\alpha} \exp(z \hat{S}) \bar{S_0}
\end{equation}
where $\bar{S_{0}} = -\hat{S} \bar{1}$. This is the distribution for one input electron. 

The distributions for $n$ input electrons can also be calculated by adding more rows to the $\hat{S}$ matrix corresponding to the zeros in the $\bar{\alpha}$ vector.
The distribution of $Z$ cannot be written in a nice closed form, since $\hat{S}$ is not diagonal in this case, but rather tridiagonal. Tools for tridiagonal matrices could perhaps be employed. For our purposes we will simplify $\hat{S}$ based on the assumption that the light levels are low. However, these ideas could possibly be used to calculate more accurate probability distributions which includes sCIC noise in a proper manner at higher light levels, thereby improving upon results, based on, say, Bayesian methods, at higher light levels. 

Assuming very low light levels, the probability that a pixel will contain an input electron from any of the sources--- parallel CIC, dark or photoelectrons--- is low. Also assuming a low probability of sCIC events, we can neglect the possibility of a sCIC event occurring in the same pixel as an input electron, and the possibility of two input electrons in the same pixel. We can then simplify the above result.
We can write a table for this setup as well:

\begin{center}
\begin{tabular}{|c|c|}
$N$ & $p$ \\
\hline 
$0$  & $1- p_p - m p_s$ \\
$X_m$ & $p_p$ \\
$X_{m-1}$ & $p_s$ \\
$\ldots$ & $\ldots$ \\
$X_1$ & $p_s$ \\
\end{tabular}
\end{center}

The first line corresponds to no photo input electrons and no sCIC electrons. The second line in the table is a photoelectron (or pCIC electron) which shows up on the input with the probability $p_p$ and no sCIC electrons, and the rest of the lines are a sCIC electron in either of the stages and no photoelectron.
This probability distribution can be described as a hyper-exponential distribution which has a simple closed form, because the $\hat{S}$ matrix in this case is diagonal.
\begin{equation} 
\hat{S} = \left( \begin{array}{cccc}
  -g_{0} & 0 & \cdots & 0 \\
  0 & -g_{m} & \cdots & 0 \\
  \vdots & \vdots & \ddots  & 0 \\
  0 & 0 & \cdots  & -g_{1}\\
  \end{array} \right) 
\end{equation}
The probability vector for this distribution is then given by 
\begin{equation}
\alpha = [ 1-p_p - mp_s, p_p, p_s, \cdots, p_s ]
\end{equation}
Now we have that 
\begin{equation} 
P \left(N = n\right) = \bar{\alpha} \exp(n \hat{S}) \bar{S_0}
\end{equation}
where $\bar{S_{0}} = -\hat{S} \bar{1}$. After taking the limit $g_{0} \rightarrow \infty$ we may simply write
\begin{eqnarray}
P(N=n) & = & (1- p_p - m p_s)\delta(n) + p_p \frac{1}{\gamma^M} \exp(-n/\gamma^m) \\ 
     & + & p_s \sum_{i=1}^{M} \frac{1}{\gamma^i} \exp(-n/\gamma^i) \nonumber
\end{eqnarray}
To arrive at the real distribution that we see in a histogram of dark frames, we need to convolve this expression with a Gaussian describing the readout noise in the readout amplifier. If we assume that the widths of the exponential distributions are much wider than the Gaussian, we can write the approximate result of the convolution as 
\begin{eqnarray}
P(n) & = & \frac{(1- p_p - m p_s)}{2\sigma \sqrt{\pi}} \exp(-n^2/(2\sigma^2)) \\
     & + & p_p \frac{1}{\gamma^m} \exp(-n/\gamma^m) \\
     & + & p_s \sum_{i=1}^{M} \frac{1}{\gamma^i} \exp(-n/\gamma^i) \nonumber
\end{eqnarray}
We can take account for a bias in the histogram in the following way
\begin{eqnarray}
\label{eq:model}
P(n) & = & \frac{(1- p_p - m p_s)}{2\sigma \sqrt{\pi}} \exp(-(n - n_0)^2/(2\sigma^2))\\
     & + & p_p\frac{1}{\gamma^m}\exp(-(n-n_0)/\gamma^m) \nonumber \\
     & + & p_s \sum_{i=1}^{m} \frac{1}{\gamma^i} \exp(-(n-n_0)/\gamma^i) \nonumber
\end{eqnarray}
where $n_0$ is the bias.